\input mtexsis
\input epsf
\paper
\doublespaced
\twelvepoint
\Eurostyletrue
\sectionminspace=0.1\vsize

\def\mea{{\int{d^4}x\sqrt{g}\,}}

\def\roc{{ \rho _{\ninepoint C}}}
\def\rov{{ \rho _{\ninepoint VAC}}}
\def\omv{{ \Omega _{\ninepoint\Lambda}}}
\def\omt{{ \Omega _{\ninepoint TOT}}}

\referencelist
\reference{price}   J.C.~Price,
               {\it Proceedings of the International Symposium on
                Experimental Gravitational Physics},
                edited by P.~Michelson, H.~En-Ke, G.~Pizzella,
              (D.~Reidel, Dordrecht, 1987)  
\endreference
\reference{kaplan}  For a recent review, see D.B.~Kaplan,
                   DOE-ER-40561-205, {\tt nucl-th/9506035}
\endreference
\reference{esm}  G.~Efstathiou, W.J.~Sutherland, and S.J.~Maddox,
                \journal Nature ;348,705 (1990)
\endreference
\reference{*esma}  M.S.~Turner, 
                \journal Phys. Scripta ;T36,167 (1991)
\endreference
\reference{*esmb}  L.M.~Krauss and M.S.~Turner, 
                CWRU-P6-95, {\tt astro-ph/9504003}
\endreference
\reference{banks}   T.~Banks,
               \journal Nucl. Phys. B;309,493 (1988); L.~Susskind, unpublished
\endreference
\reference{*banksa} See also, S.R.~Beane, 
                \journal Phys. Lett. B;358,203 (1995)
\endreference
\reference{jd}  See, for example, J.F~Donoghue,
                 \journal Phys. Rev. D;50,3874 (1994)
\endreference
\reference{sw}  S.~Weinberg, \journal Rev. Mod. Phy.;61,1 (1989)
\endreference
\reference{*swa}  S.M.~Carroll and W.H.~Press, 
         \journal Annu. Rev. Astron. Astrophys.;30,499 (1992)
\endreference
\reference{giu}  See, for example, S.~Dimopoulos and G.F.~Giudice, 
              CERN-TH/96-47, {\tt hep-ph/9602350}
\endreference
\reference{spar} M.J.~Sparnaay, 
                 \journal Physica;24,751 (1958)
\endreference
\reference{hos}  J.K.~Hoskins {\it et al},
                 \journal Phys. Rev. D;32,3084 (1985)
\endreference
\endreferencelist

\titlepage
\hskip5in{DUKE-TH-96-108}

\hskip5in{hep-ph/9702419}
\title
On the importance of testing gravity
at distances less than 1cm
\endtitle
\author
Silas R.~Beane\Footnote\dag{Present Address: Physics Department, University of Maryland, College Park, MD 20742}
Department of Physics, Duke University
Durham, NC 27708-0305
\endauthor
\abstract
\doublespaced
If the mechanism responsible for the smallness of the vacuum energy is
consistent with local quantum field theory, general arguments suggest
the existence of at least one unobserved scalar particle with Compton
wavelength bounded from below by one tenth of a millimeter. We show
that this bound is saturated if vacuum energy is a substantial
component of the energy density of the universe. Therefore, the
success of cosmological models with a significant vacuum energy
component suggests the existence of new macroscopic forces with range
in the sub-millimeter region. There are virtually no experimental
constraints on the existence of quanta with this range of interaction.
\endabstract 
\endtitlepage
\vfill\eject                                     
\doublespaced
\superrefsfalse

There are no significant experimental constraints on new
gravitational-strength forces at distances shorter than 1 cm.  The
present situation is illustrated in Figure 1: a plot of wavelength
($\lambda$) vs. the strength of a weak Yukawa force relative to
gravity ($\alpha$)\ref{price}.  In order to identify the
characteristic scale in question, note that the dashed vertical lines
cutting through the figure correspond to the Compton wavelength
associated with the critical energy density of the universe.  The
critical energy density of the universe is given by $\roc
=3{H_0^2}/8\pi{G_N}=(3.0{\sqrt h}\times{10^{-3}}\;{\rm eV})^4$, where
Hubble's constant, $H_0=100h\; {\rm km\; s^{-1}\;Mpc^{-1}}$, is
consistent with observations if $h=0.4-0.9$. {\it A priori} the
overlap of this characteristic scale with the experimentally
unexplored region in Figure 1 is not particularly interesting.
However, in this essay we will argue that if a substantial fraction of
the energy density of the universe is in the form of vacuum energy,
then the overlap of the dashed lines with the experimentally
unexplored region of Figure 1 becomes a coincidence of fundamental
interest. Our argument is based on two assumptions:
\vskip0.15in

\noindent (A) {\it Local quantum field theory always works.}
\vskip0.15in

\noindent (B) {\it 
In the present epoch a substantial component of the
\hfill\break\hphantom{ijk}\hskip0.15in energy density of the universe
is vacuum energy.}
\vskip0.15in

\noindent Several comments are in order. Assumption (A) is
simply the statement that {\it all} experimentally accessible physical
phenomena can be described using effective quantum field
theories\ref{kaplan}.  The motivation for (B) is well known;
cosmological models with $\rov\sim\roc$ have many attractive
features\ref{esm}. In particular, cold dark matter models with $\omv
={\rov /\roc}\simeq 0.65$ $({\omt}=1)$ are consistent with a wide
variety of observations; for example, vacuum energy of this magnitude
settles the age paradox and allows for a flat universe without
contradicting measures of matter density\ref{esm}.  We will argue that
if (A) is not violated, then the smallness of the cosmological
constant implies the existence of quanta not yet seen
experimentally\ref{banks}.  This argument ---which we call the
Banks-Susskind theorem--- is powerful since there is {\it no evidence
whatsoever} that (A) is violated in nature.  We will argue on the
basis of this theorem that if (B) is true, then the range of these
unobserved quanta should fall in the experimentally unexplored region
of Figure 1.

Effective quantum field theories ---the current paradigm in particle
physics--- imply the existence of many disparate contributions to the
vacuum energy ---from zero-point fluctuations of the electromagnetic
field to non-perturbative phenomena like spontaneous breaking of
chiral symmetry in quantum chromodynamics.  Here we will be concerned
with the vacuum energy which is relevant at late times and large
distance scales and which therefore plays a part in the evolution of
the universe in the present epoch. This vacuum energy is constrained
experimentally and can be defined in the context of the relevant
effective quantum field theory. Since the relevant scales are
macroscopic, the effective theory we are interested in has all massive
particles in nature integrated out (see Figure 2). The lightest
massive particle is the electron, and so the cutoff of the effective
theory can be taken as the electron mass. This effective theory
encodes the interactions of photons, gravitons and neutrinos in a
manner consistent with the assumed symmetries. Here for simplicity we
will neglect the neutrino and photon interactions ---except for their
possible contributions to the vacuum energy.

The action of gravity is determined by invariance under
general coordinate transformations:

$${S}[g]=
\mea\lbrack{-{\Lambda}-{{M_p^2}\over{16\pi}}R+{\alpha}{R^2}
+{\beta}{R_{\mu\nu}R^{\mu\nu}} +O{(\partial ^6)}}\rbrack, 
\EQN grav1 $$ 
where $g\equiv -detg_{\mu\nu}$, $R_{\mu\nu}$ is the Ricci tensor, and
$R$ is the curvature scalar.  The last term denotes invariant
contributions with six or more derivatives of the metric field.
Experiment determines $M_p\simeq{10^{19}}GeV$.  There are experimental
bounds on $\alpha$ and $\beta$\ref{jd}: $|\alpha|,|\beta|\leq
{10^{74}}$. These bounds are weak because these terms are suppressed
by powers of ${q/{M_p}}$ where $q$ is a characteristic momentum in the
low-energy effective theory.  A rough observational bound on
$\Lambda$, the vacuum energy or cosmological constant, is\ref{sw}

$$|\,\Lambda\,|\leq{\roc}\simeq 10^{-47} GeV^4. \EQN grav8 $$ The
cosmological constant problem is the simple fact that $\Lambda$ is
much smaller than any characteristic mass scale in elementary particle
physics, whereas basic field theory dimensional analysis does not rule
out $\Lambda\sim{M_p^4}\simeq 10^{74} GeV^4$.  Perhaps more
discouraging than the size of the discrepancy is the number of
disparate contributions which evidently sum to a small number.

What about the value of $\Lambda$ in the macroscopic effective theory?
Consider the ladder of effective field theories illustrated in Figure
2.  Each rung of the ladder corresponds to a threshold at which a
massive particle is integrated out or some non-perturbative phenomenon
takes place. In the macroscopic effective theory (region ${\rm III}$)
we can write the total cosmological constant schematically as

$${\Lambda}={\Lambda^{{\rm III}}}+{\Lambda^{{\rm II}}}+{\Lambda^{{\rm
I}}} +...  \EQN towercontr$$ Since there is no symmetry to forbid it,
we would expect that zero-point fluctuations of quantum fields in each
effective theory give a contribution to $\Lambda$ of order the
momentum cutoff.  On purely dimensional grounds ---ignoring
geometrical factors of $2$ and $\pi$--- we expect

$${\Lambda^{{\rm III}}}\sim{m_e^4}\simeq 10^{-13} GeV^4,
\EQN econtr$$ 
which is 32 orders of magnitude larger than the upper bound, and  

$${\Lambda^{{\rm II}}}\sim{m_\mu^4}\simeq 10^{-4} GeV^4,
\EQN mucontr$$ 
which is 43 orders of magnitude larger than the upper bound.  Even if
we assume that the effective cosmological constant in region $\rm I$
vanishes, we have a severe cosmological constant problem in the
macroscopic effective theory.  We can make $\Lambda^{{\rm II}}$ and
$\Lambda^{\rm III}$ cancel by introducing an arbitrary coefficient
tuned to one place in a billion. One might think that a symmetry is
capable of explaining this sort of correlation. However, symmetry
generators which act locally on the fields carry no energy and
momentum and cannot relate the vacuum energy associated with distinct
effective field theories\ref{banks}.  Evidently, the only way around
this impasse which is consistent with local quantum field theory is to
acknowledge the existence of quanta which have not been seen
experimentally, and which have therefore been inadvertently left out
of the effective theory description.  This is the essence of the
Banks-Susskind theorem\ref{banks}.  We denote this field or fields
collectively as $\phi$. Presumably $\phi$ carries vacuum quantum
numbers as it must by some means act as a source of the
energy-momentum tensor in the vacuum\ref{sw}. It is important to
stress that no satisfactory mechanism involving $\phi$ has been
found\ref{sw}. Here we have argued that if this {\it unknown}
mechanism is consistent with local quantum field theory, then $\phi$
should be a fundamental ingredient.

In this picture the cosmological constant relevant to cosmology is an
effect arising from the decoupling of $\phi$.  That is, at distance
scales shorter than $\phi$'s Compton wavelength, the dynamics of
$\phi$, by assumption, ensure a vanishing cosmological constant.  On
the other hand, when probing distances greater than $\phi$'s Compton
wavelength, $\phi$ gets integrated out and there is no longer a
mechanism to prevent gravitational and eletromagnetic fluctuations in
the vacuum.  Hence at these distance scales we expect

$$\Lambda\sim {m_\phi^4}, \EQN asdk$$ which together with \Eq{grav8}
implies that ${m_\phi}\leq 3.0{\sqrt{h}}\times{10^{-3}}\;{\rm eV}$,
with associated Compton wavelength $\lambda_\phi\geq 6.6{(h)^{-{1\over
2}}}\times{10^{-5}}\;{\rm m}$.  There is an interesting consequence of
this scenario for $\Lambda$.  The only {\it strictly} massless
particles in nature are associated with gauge invariance or general
coordinate invariance and therefore transform as vectors and tensors
under the Lorentz group.  There are no strictly massless scalars. The
only natural light scalars are Goldstone bosons, which arise from the
spontaneous breakdown of global symmetries. Such symmetries are not
exact in nature, and so in this conservative picture $\Lambda $
necessarily takes a non-zero value.

On the basis of assumption (B) we give the conservative lower bound
$|\,\Lambda\,|\geq{(0.1)}{\roc}$, which in turn bounds the mass of
$\phi$:

$$1.7{\sqrt{h}}\times{10^{-3}}\;{\rm eV}\leq {m_\phi}\leq
3.0{\sqrt{h}}\times{10^{-3}}\;{\rm eV}. \EQN fullbo$$ This places
$\phi$'s range around the dashed lines in the experimentally
unexplored region of the parameter space illustrated in Figure 1.  Of
course the coupling strength of $\phi$ to matter is also a relevant
parameter.  In order that $\phi$ eliminate vacuum fluctuations up to
the Planck scale, $\phi$ must originate at the Planck scale. Hence it
is natural that $\phi$ couple weakly to matter; if we assume the
simple Yukawa form:

$$g{{m_N}\over{M_p}}\phi{\bar N}N \EQN yuk$$ 
where $N$ is the nucleon field, we obtain
${g^2}/{4\pi}=\alpha$\ref{giu}, where $\alpha$ is the usual coupling
parameter\ref{price}.  On the basis of naive dimensional analysis $g$
---and therefore $\alpha$--- is expected to be of order one. However,
the coupling strength can vary substantially with detailed dynamical
assumptions\ref{giu}.

What is the present status of experimental searches for new
gravitational strength forces in the sub-cm region?  Existing limits
are illustrated in Figure 1.  The curve labelled Sparnaay is deduced
from a classic electromagnetic Casimir force measurement\ref{spar}.
This experiment measured the attractive force between parallel plates
at separations of roughly $10^{-1}$ $\mu m$ to $10$ $\mu m$. Bounds on
$\alpha$ were extracted from Sparnaay data by conjecturing a force due
to a Yukawa interaction between parallel plates (see \Ref{price}).
The curve labelled Hoskins {\it et. al.} is deduced from the
Cavendish-type experiment of \Ref{hos}. This experiment searched for
deviations from the inverse-square law in the $2-5$ $cm$ region. The
bounds on $\alpha$ are an extrapolation of these results to shorter
distances.  It is clear that there are no significant bounds in the
sub-cm region.

Cryogenic mechanical oscillator techniques have been
proposed\ref{price} which would improve existing limits on the
strength of a Yukawa force with a range of $100$ $\mu$m by up to
$10^{10}$.  This is precisely the range which our theoretical argument
finds most interesting.  The dotted curve in Figure 1 indicates the
sensitivity of this experiment. The important background effects are
due to: vibrations generated by the motion of the source mass,
Newtonian background due to edge effects and geometry defects, and
magnetic and electrostatic forces\ref{price}. The analysis leading to
the dotted curve is given in \Ref{price}. There also exists
independent theoretical motivation for probing this region; it has
recently been argued\ref{giu} that masses and couplings of scalar
fields which arise in certain classes of supersymmetric theories fall
naturally into the region of parameter space accessible to the
cryogenic oscillator.

In summary, if local field theory always works, the observational fact
that the cosmological constant is small implies the existence of
quanta that have not been observed. This is probably the most
conservative statement one can make about the cosmological constant
problem. We have argued on the basis of this ``theorem'' that a small
non-vanishing cosmological constant of order the critical energy
density of the universe suggests the existence of new macroscopic
forces in the 100 $\mu$m region.  On the basis of general physical
principles and established cosmological observations, we hope to have
convinced the reader that experimental tests of the inverse-square law
in the sub-cm range should be vigorously pursued.

\vskip0.25in

\noindent This work was supported by the
U.S. Department of Energy (Grant DE-FG05-90ER40592).  I thank
S.~Matinyan, B.~M{\" u}ller, T.J.~Phillips and R.P.~Springer for
constructive criticism. This essay received honorable mention in the
1996 {\it Gravity Research Foundation} essay competition.

\nosechead{References}
\ListReferences
\vfill\supereject

\midfigure{price}
\epsfxsize 6in
\centerline{\epsfbox{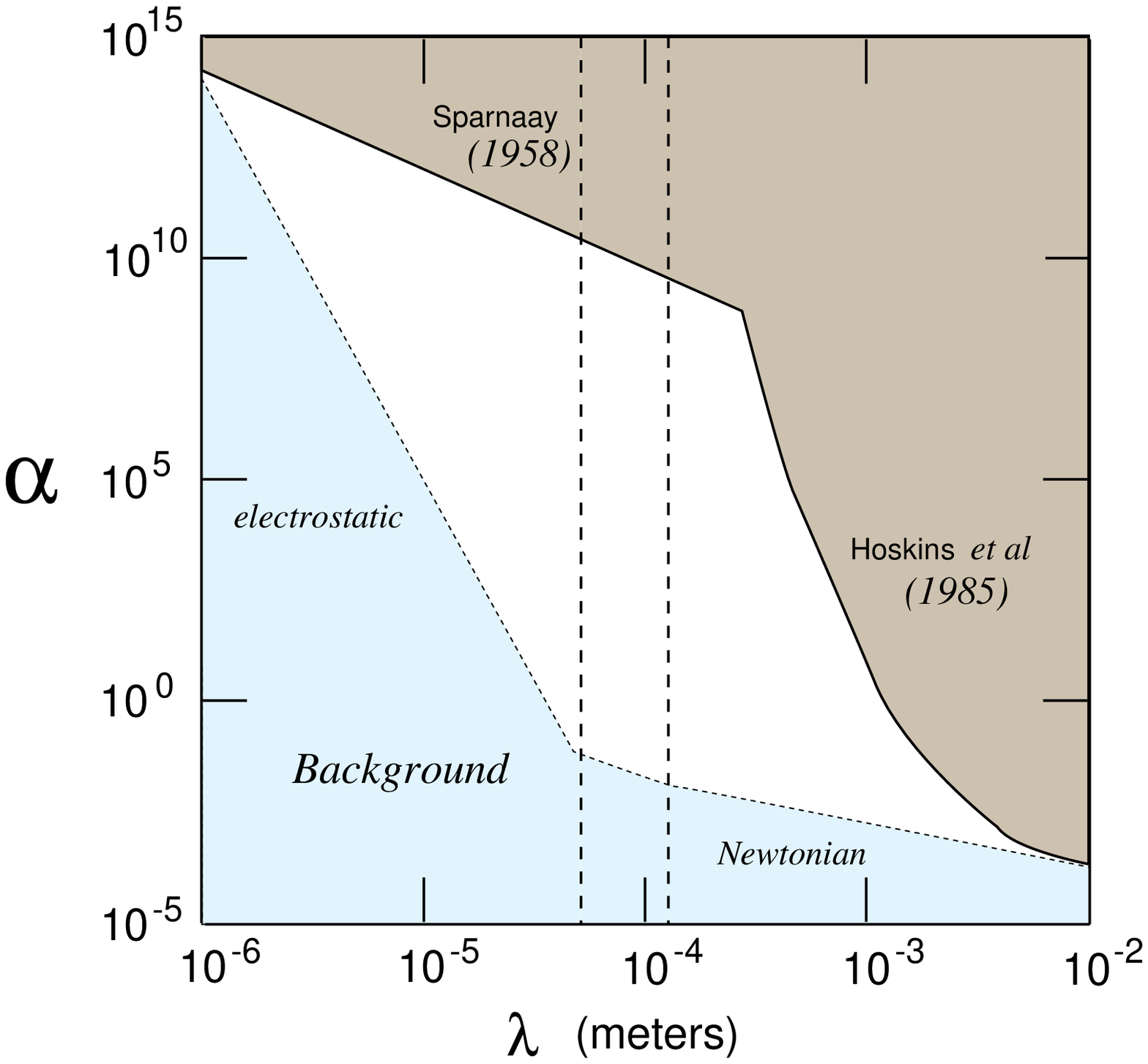}}
\caption{\doublespaced Wavelength ($\lambda$) vs. 
the strength of a weak Yukawa force
relative to gravity ($\alpha$). The region above the solid line is
excluded by Sparnaay (electromagnetic Casimir force
measurements\ref{spar}) and by Hoskins {\it et al} (Cavendish-type
experiment\ref{hos}).  The region between the solid line and the
dotted line is accessible to the cryogenic mechanical
oscillator\ref{price}.  The region below the dotted line is
inaccessible due to Newtonian and electrostatic backgrounds.  The
dashed lines represent the Compton wavelength associated with the
critical energy density of the universe for $h=0.4$ (right) and
$h=0.9$ (left).}
\endfigure
\vfill\eject

\midfigure{efftheory}
\epsfxsize 6in
\centerline{\epsfbox{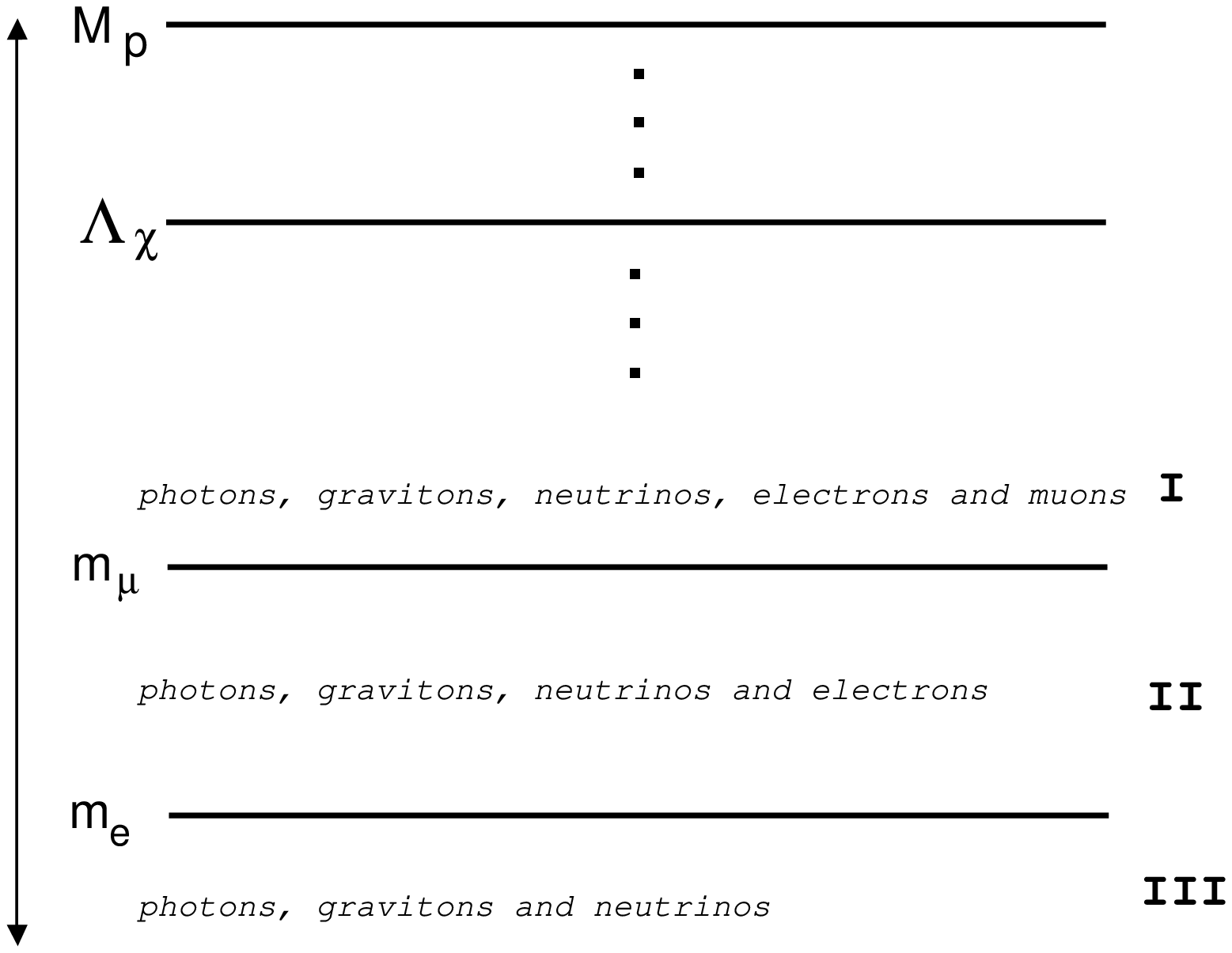}}
\caption{\doublespaced The ladder of effective field theories.
All massive particles in nature are represented by a rung on the
ladder.  The chiral symmetry breaking scale, $\Lambda_\chi$, is an
example of a rung on the ladder arising from non-perturbative
phenomena.  Here we are interested in the lowest rungs.
Region ${\rm III}$ is the effective theory with all massive particles
integrated out and is therefore relevant to macroscopic phenomena.}
\endfigure

\end